%% file: 0-main.tex
\documentclass[10pt,conference]{IEEEtran}

\usepackage{algorithm}
\usepackage{amsmath}
\usepackage{array}
\usepackage{booktabs}
\usepackage[font=small,labelfont=bf]{caption}
\usepackage{color}
\usepackage{colortbl}
\usepackage{enumitem}
\usepackage{graphicx}
\usepackage{hyperref}
\usepackage[capitalise]{cleveref}
\usepackage{listings}
\usepackage{mdframed}
\usepackage{MnSymbol}
\usepackage{multirow}
\usepackage{soul}
\usepackage{subcaption}
\usepackage{tabularx}
\usepackage{tcolorbox}
\usepackage{tikz}
\usepackage{titlesec}
\usepackage{xurl}
\usepackage{wrapfig}
\usepackage{xcolor}
\usepackage{balance}

\definecolor{latex_green}{HTML}{4c886b}
\definecolor{latex_cyan}{HTML}{318495}
\definecolor{latex_blue}{HTML}{0000ff}
\definecolor{latex_gray}{HTML}{646464}
\definecolor{latex_purple}{HTML}{7d2884}

\lstdefinestyle{lststyle}{
    commentstyle=\color{latex_green},
    keywordstyle=\color{latex_blue},
    numberstyle=\tiny\color{latex_gray},
    stringstyle=\color{latex_purple},
    basicstyle=\ttfamily\footnotesize,
    breakatwhitespace=false,         
    breaklines=true,                 
    captionpos=b,                    
    keepspaces=true,                 
    numbers=left,                    
    numbersep=5pt,                  
    showspaces=false,                
    showstringspaces=false,
    showtabs=false,                  
    tabsize=2
}
\newcommand{\blacktextcircled}[1]{\tikz[baseline=(char.base)] \node[shape=circle,fill=black,text=white,inner sep=1pt] (char) {#1};}
\newcommand{\whitetextcircled}[1]{\tikz[baseline=(char.base)] \node[shape=circle,draw=black,fill=white,text=black,inner sep=1pt] (char) {#1};}
\newcolumntype{Y}{>{\centering\arraybackslash}X}
\newcommand{\midcomment}[1]{%
}

\makeatletter
\def\subsubsection{
  \@startsection{subsubsection}
  {3}
  {\parindent}
  {0ex plus 0.1ex minus 0.1ex}
  {0ex}
  {\normalfont\normalsize\itshape}
}
\newcommand{\linebreakand}{%
  \end{@IEEEauthorhalign}
  \hfill\mbox{}\par
  \mbox{}\hfill\begin{@IEEEauthorhalign}
}
\makeatother

\begin{document}

\title{\texttt{DynaCI}: Generating GitHub Actions Workflows via Dynamic Analysis}

\author{
    \IEEEauthorblockN{1\textsuperscript{st} Javid Ditty}
    \IEEEauthorblockA{\textit{University of Michigan-Dearborn}\\
    Dearborn, Michigan, USA \\
    jditty@umich.edu}
    \and
    \IEEEauthorblockN{2\textsuperscript{nd} Rahul Ghanshyam Chavan}
    \IEEEauthorblockA{\textit{University of Michigan-Dearborn}\\
    Dearborn, Michigan, USA \\
    rchavan@umich.edu}
    \linebreakand
    \IEEEauthorblockN{3\textsuperscript{rd} Nimmi Weeraddana}
    \IEEEauthorblockA{\textit{University of Calgary}\\
    Calgary, Alberta, Canada \\
    nimmi.weeraddana@ucalgary.ca}
    \and
    \IEEEauthorblockN{4\textsuperscript{th} Foyzul Hassan}
    \IEEEauthorblockA{\textit{University of Michigan-Dearborn}\\
    Dearborn, Michigan, USA \\
    foyzul@umich.edu}
}

\maketitle

\begin{abstract}
Continuous Integration (CI) is an essential practice that automatically validates code changes through builds and tests. However, configuring CI workflows remains challenging, as modern software often features complex architectures, numerous dependencies, and support for diverse execution environments. Most CI platforms also require workflows to be written in their particular domain-specific languages. As a result, CI workflows are often error-prone and hard to maintain, especially for those unfamiliar with CI semantics. Prior work on automated CI workflow generation require developers to create specifications that describe their intended workflows using natural language, partial workflows, etc. However, in practice, developers may struggle to articulate requirements and may be unaware of available CI features.

To address these limitations, we propose \texttt{DynaCI}, a novel approach that uses runtime information from developers' local environments to generate CI workflows. \texttt{DynaCI} (i) traces the developers' local environment to collect runtime information, (ii) constructs an initial workflow by applying runtime-informed heuristics, and (iii) iteratively refines the workflow via LLM-based Retrieval Augmented Generation (RAG). We evaluate \texttt{DynaCI} on 50 popular Python projects that use GitHub Actions, a popular CI platform, and compare it against the state-of-the-art CI workflow generation approaches: \texttt{GH-WCOM} and \texttt{LLM4GW}. \texttt{DynaCI} outperforms these existing approaches by more than 21.13\% in BLEU score and 20.96\% in F1 score on average. It also generates workflows without missing functionality in 56\% of cases, while requiring fewer manual modifications to repair such exceptions. These results demonstrate that leveraging runtime information substantially improves LLM-based CI workflow generation, and offer a promising direction for broader DevOps configuration automation.
\end{abstract}


\input{1-Introduction}
\input{2-Background}
\input{3-Approach}
\input{4-Evaluation}
\input{5-Threats}
\input{6-Related}
\input{7-Conclusion}

\newpage
\balance
\bibliographystyle{IEEEtran}
\bibliography{biblio}

\end{document}

%% file: 1-Introduction.tex
\section{Introduction}
\label{sec:introduction}

Continuous Integration (CI) is a popular software engineering practice that automates the code integration process in software repositories~\cite{Vasilescu2015}. 
Prior work shows that CI enables faster pull request acceptance and frequent releases~\cite{hilton_2016_ci}, increases the number of change merges~\cite{zhao_2017_ci}, and boosts developer productivity, efficiency, and confidence~\cite{soares_2022_ci}.

While there are many CI platforms~\cite{travis_ci, jenkins_ci, gitlab_ci}, GitHub Actions (GHA) remains the most widely used~\cite{hilton_2016_ci, golzadeh_2022_ci}, with at least 71.8\% of popular GitHub repositories containing a \textit{workflow} since 2023 \cite{decan_2022_gha, cardoen_2024_gha}. Since December 2025, developers can define a GHA workflow in a YAML using 84 different components, such as event triggers, jobs, steps, actions, runners, matrices, containers, secrets, and conditional rules~\cite{github_2025_workflows}. With such a capacity for extensive customization, configuring CI workflows is inherently challenging for developers. Saroar et al.~\cite{saroar_2023_gha} surveyed 90 developers and found that 60.87\% often encounter difficulties writing workflows. Among these, the most common challenges were related to confusing and error-prone syntax. Zhang et al.~\cite{zhang_2024_ci} studied 6,590 Stack Overflow questions and 315 GitHub issues, and found that workflow configuration issues were discussed more than those related to dependencies, security, integration, deployment, etc. Again, the most common problem among these issues was workflows' "nightmare" syntax~\cite{zhang_2024_ci}. Valenzuela-Toledo et al.~\cite{valenzuela-toledo_2022_gha} has argued that this trend is due, in part, to inadequate tooling support for writing GHA workflows.

Previous works have attempted to provide this tooling support. For example, Mastropaolo et al.~\cite{mastropaolo_2024_synthesis} propose \texttt{GH-WCOM}, a transformer-based approach that provides code completion suggestions for GHA workflows. Zhang et al.~\cite{zhang_2024_synthesis} propose \texttt{LLM4GW}, a set of prompting techniques that optimize the performance of Large Language Models (LLMs) on GHA workflow generation, syntactic error identification, code injection vulnerability detection, syntactic error fixing, and code injection vulnerability repair. However, despite their intended benefits, these tools may fall short in practice, as they still require inputs that demand substantial domain knowledge of GHA to produce acceptable configurations. \texttt{GH-WCOM}~\cite{mastropaolo_2024_synthesis}, for example, still requires a developer to write the majority of their workflows beforehand so that it can fill in any gaps. \texttt{LLM4GW}~\cite{zhang_2024_synthesis} accepts natural language instructions instead, but developers must provide detailed workflow-, job-, and step-level specifications for functional correctness.

To address these concerns, we propose \texttt{DynaCI}, the first tool, to the best of our knowledge, that combines runtime analysis with LLMs to synthesize GHA workflows. Our intuition is that a developer’s local environment can serve as a starting point for extracting the information required for generating CI workflows. As the developer runs test suites in their local environment, \texttt{DynaCI} collects runtime traces and extracts information about the executed commands, dependencies, and execution order. 
It then applies heuristics to select candidate GHA settings (i.e., components), assembles an initial workflow, and iteratively refines it using LLM-based Retrieval Augmented Generation (RAG)~\cite{fan_2024_rag, arslan2024survey}.
We evaluate \texttt{DynaCI} against \texttt{GH-WCOM} and \texttt{LLM4GW}, the state-of-the-art in GHA workflow synthesis tools, on 50 popular Python projects to answer the following research questions:

$\mathbf{RQ_{1}}$ \textbf{How effective is \texttt{DynaCI} at generating CI workflows?} To assess whether \texttt{DynaCI} outperforms the state-of-the-art, we compare the tools' workflows against their targets across several popular workflow similarity metrics. We found that \texttt{DynaCI} has a BLEU score~\cite{papineni_2002_metric} of 52.81\% and a F1 score of 62.06\%, outperforming the state-of-the-art by more than 21.13\% and 20.96\%, respectively. 

$\mathbf{RQ_{2}}$ \textbf{How time and cost efficient is \texttt{DynaCI}?} Since \texttt{DynaCI} uses an LLM to iteratively refine workflows, we measure its overhead in terms of elapsed time and LLM token usage per workflow. We found that \texttt{DynaCI} incurs a median overhead of 5.55 minutes and 653,755 tokens per workflow. 

$\mathbf{RQ_{3}}$ \textbf{How much modification is needed for CI workflows generated by \texttt{DynaCI}?} Like other workflow synthesis tools, \texttt{DynaCI} may generate workflows that lack intended functionality. In our manual analysis, we found that only 44\% of \texttt{DynaCI} workflows lacked intended functionality, whereas \textit{every} workflow from state-of-the-art tools needed modification.

\vspace{0.5em}
\noindent This paper makes the following contributions:
\begin{itemize}[leftmargin=1em]
    \item It presents \texttt{DynaCI}, the first approach, to the best of our knowledge, that synthesizes GHA workflows using runtime information and LLMs;
    \item It conducts an empirical analysis of \texttt{DynaCI} on 50 popular, Python-based GitHub projects, showing that it is significantly more effective than existing tools with little overhead and need for modification;
    \item It conducts an ablation study, showing that \texttt{DynaCI}'s hybrid approach plays a substantial role beyond what can be attributed to its heuristics or the LLM alone.
\end{itemize}

%% file: 2-Background.tex
\section{Background and Motivation}
\label{sec:background}

\subsection{GitHub Actions Workflows}
\label{subsec:background-workflows}
\begin{figure}[t]
    \centering
    \includegraphics[width=\linewidth]{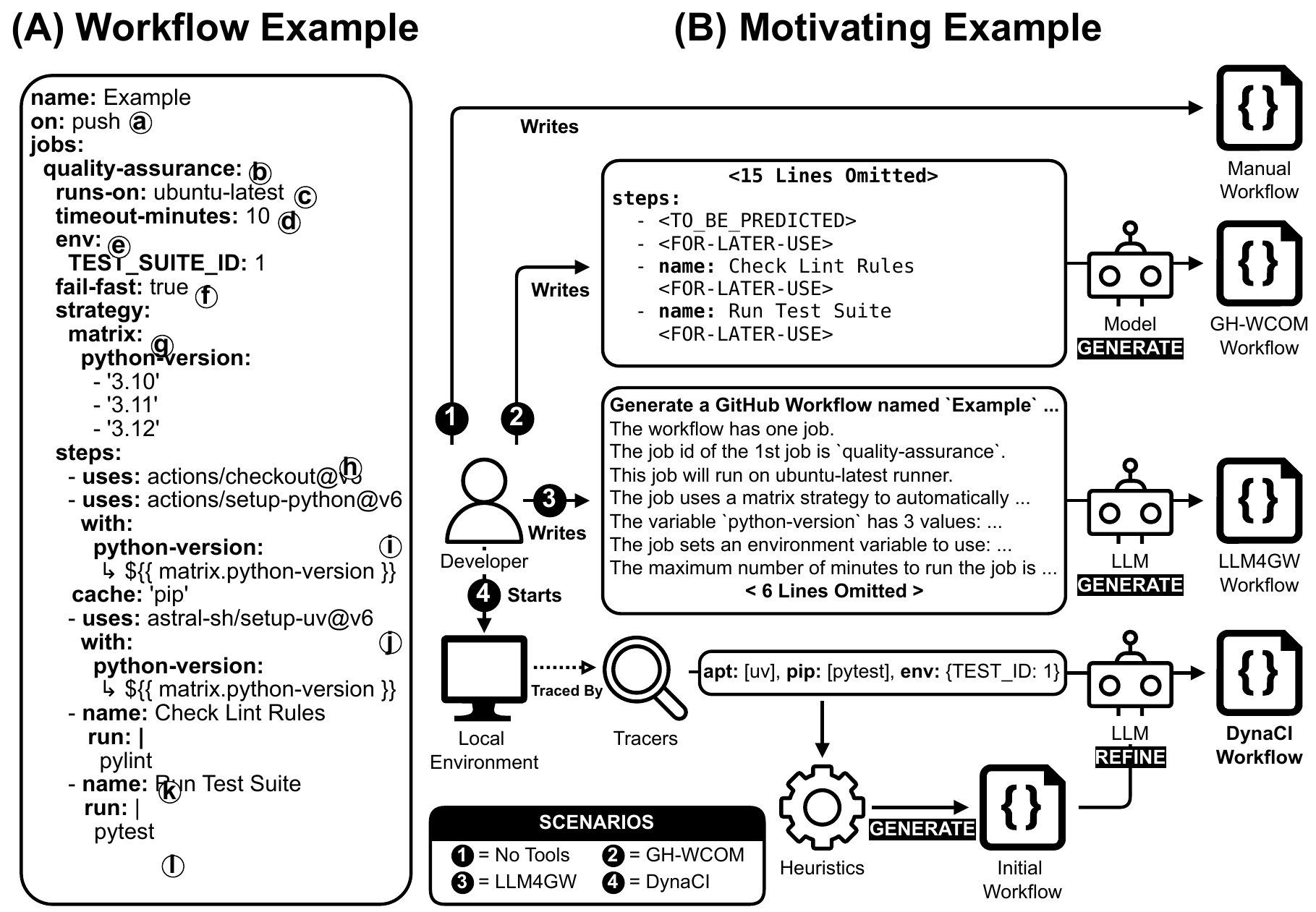}
    \caption{Comparing GHA Workflow Synthesis Across Four Scenarios}
    \label{fig:background-example}
    \vspace{-0.5cm}
\end{figure}

GHA workflows automate the building, testing, and deployment activities for a GitHub repository and are configured through YAML files. \autoref{fig:background-example} (A) shows an example configuration file that illustrates the main components of a GHA workflow. A workflow is triggered in response to an event \whitetextcircled{a}, which executes its jobs. A job \whitetextcircled{b} is a sequence of steps that run on a shared machine \whitetextcircled{c} to accomplish a set of tasks. It can initialize environmental variables \whitetextcircled{e}, set a maximum number of minutes to run \whitetextcircled{d}, use a matrix that repeats the job with different configurations \whitetextcircled{g}, and stop queued configurations if any fails \whitetextcircled{f}. A step is a group of shell commands that accomplish a single task, such as \whitetextcircled{k} or \whitetextcircled{l}, which performs software quality assurance. An action is a reusable step whose contents are hosted in an external repository. For example, \whitetextcircled{h} checks out a repository into the runner so that the workflow can access it, \whitetextcircled{i} sets up a particular Python version in the execution environment, and \whitetextcircled{k} does the same with \texttt{uv}, a popular Python package manager. These, alongside services, containers, and 84 additional settings \cite{github_2025_syntax}, make workflows difficult to write without supporting tools~\cite{saroar_2023_gha, zhang_2024_ci, valenzuela-toledo_2022_gha}.

\subsection{Motivating Example}
\label{subsec:background-example}
\autoref{fig:background-example} (B) presents four scenarios in which a developer tries to create a workflow. In  scenario \blacktextcircled{1}, which is the most common, the developer attempts to write the entire workflow without using tools. This requires them to (i) articulate their needs and (ii) leverage substantial domain knowledge about GHA. Since most developers do not meet either condition \cite{saroar_2023_gha},  scenario \blacktextcircled{2} introduces \texttt{GH-WCOM}~\cite{mastropaolo_2024_synthesis}, a tool that auto-completes any gaps that a developer leaves behind in a GHA workflow file. While this mitigates the knowledge requirement, the developer must still be familiar enough with GHA to write a partial workflow. For those who are not familiar with writing GHA configuration files, \texttt{LLM4GW}~\cite{zhang_2024_synthesis} can be used, as shown in scenario \blacktextcircled{3}, where the developer writes natural language instructions instead of a (partial) workflow. However, the instructions required for this scenario can still be complex and require substantial domain knowledge about GHA, such that developers may expend effort comparable to that required in the first scenario. To fill the gaps in the previous scenarios, we introduce \texttt{DynaCI}---scenario \blacktextcircled{4} in \autoref{fig:background-example} (B). \texttt{DynaCI} traces the developer’s local environment during execution and uses the runtime information to generate and refine the workflow. 
Besides performing their regular development and testing tasks and attaching \texttt{DynaCI} to their environment, the developer is not required to produce additional inputs, such as workflow components (e.g., \blacktextcircled{1} and \blacktextcircled{2}) or natural language instructions~(e.g., \blacktextcircled{3}).

%% file: 3-Approach.tex
\section{Approach}
\label{sec:approach}
\begin{figure}[t]
    \centering
    \includegraphics[width=0.9\linewidth]{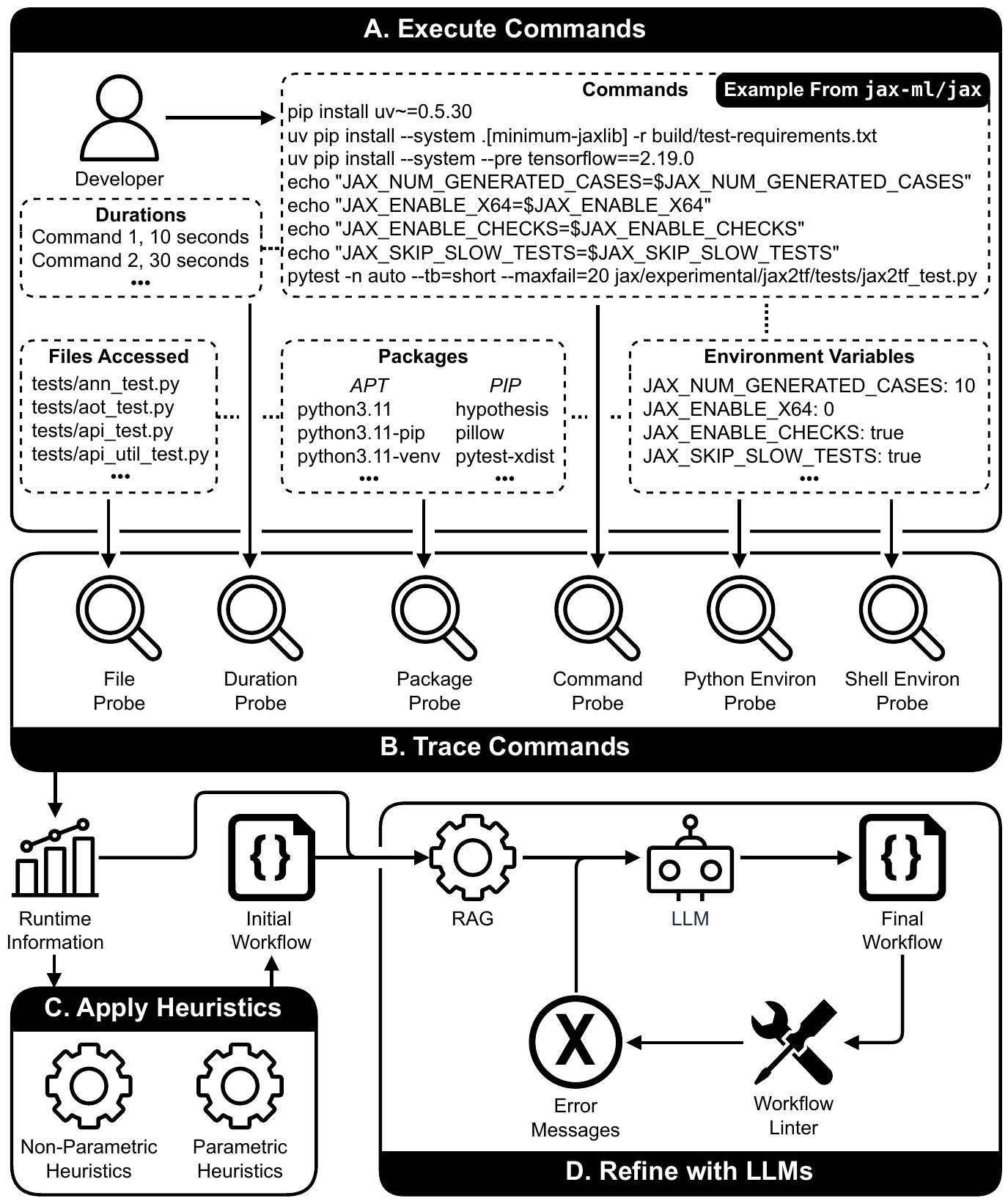}
    \caption[Overview of \texttt{DynaCI}]{Overview of \texttt{DynaCI}}
    \label{fig:overview_dynaci}
\end{figure}

\noindent This section presents a detailed overview of \texttt{DynaCI}, and \autoref{fig:overview_dynaci} shows its main steps: (A) Execute Commands, (B) Trace Commands, (C) Apply Heuristics, and (D) Refine with LLMs. 
Together, these steps produce a GitHub workflow that reproduces the developer's local environment.
We designed \texttt{DynaCI} to handle Python environments, since it is a popular language in projects that use workflows~\cite{decan_2022_gha}, but our approach could be generalized to other languages, such as Java. 

\subsection{Execute Commands}
\label{subsec:approach-execute}
\vspace{-0.1cm}
\begin{figure}[t]
    \centering
    \includegraphics[width=0.7\linewidth]{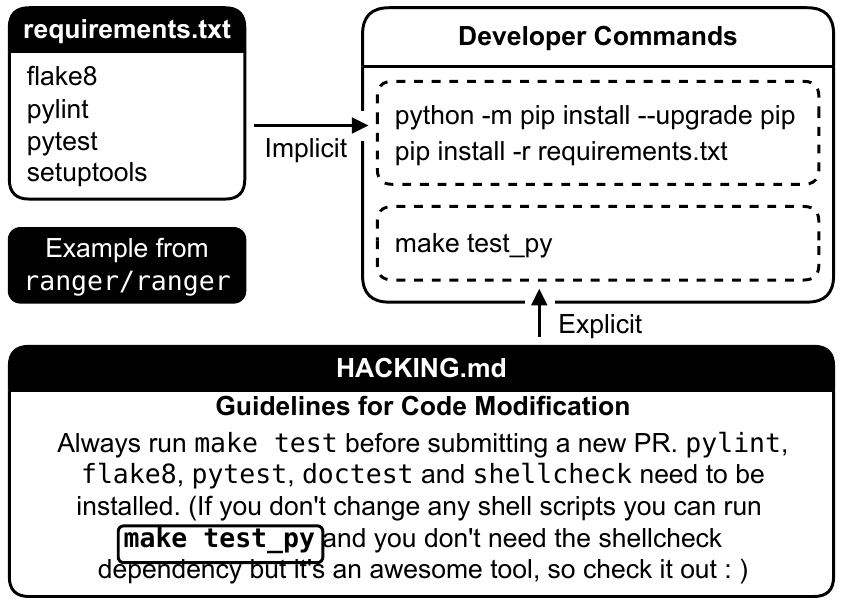}
    \caption[Example of Manually Identifying Developer Commands]{Example of Manually Identifying Developer Commands}
    \label{fig:overview_local}
    \vspace{-0.4cm}
\end{figure}

Software testing is crucial for improving code quality and reliability~\cite{chen_2017_testing}. Before contributing code changes to shared repositories, developers typically validate their changes in local development environments, where test executions have access to project-specific resources such as test files, software packages, and environment variables~\cite{duvall2007ci}. To perform these validation activities, developers execute build and test commands that exercise the project’s functionality. This fundamental software engineering practice helps detect faults before code changes are pushed to shared repositories, avoiding unnecessary CI wait times~\cite{rostami_2023_ci}. 

Developers usually record their build and test commands in project documentation (e.g., \texttt{README.md}, \texttt{CONTRIBUTING.md}, \texttt{HACKING.md}, and webpages) and configurations (e.g., \texttt{pyproject.toml}, \texttt{setup.py}) to help new code contributors implement the practice in their own local environments. 
\autoref{fig:overview_local} provides an example of this in the \texttt{ranger/ranger} project. In this example, the \texttt{HACKING.md} file provides \textit{explicit} guidance on how to build and test the project, e.g., run \texttt{make test\_py} to only test Python scripts. However, these stated command(s) \textit{implicitly} depend on Python dependencies that must be installed (via pip commands) from the \texttt{requirements.txt} file.

\vspace{-0.5em}
\subsection{Trace Commands}
\label{subsec:approach-trace}
Once a developer identifies their project's building and testing commands (via its documentation and configurations), they can run those commands in their local environment.
To start generating a CI workflow, the developer provides their set of commands, as an input file, to \texttt{DynaCI}, which will execute them in the same environment \textit{twice} to trace their runtime information. 
In the first execution (\autoref{subsubsec:trace-first}), it captures runtime traces, such as durations of commands, file accesses in system calls, and dependency installations in package managers, using \texttt{strace}~\cite{kranenburg_1991_strace}. 
In the second execution (\autoref{subsubsec:trace-second}), \texttt{DynaCI} captures environment variable accesses using \texttt{ltrace}~\cite{cespedes_2019_ltrace}. 
This separation is necessary since multiple \texttt{ptrace} tools, such as \texttt{strace} and \texttt{ltrace}, cannot attach to the same process at the same time~\cite{manual_2025_EPERM}. 
Below, we discuss these two execution runs in detail.

\subsubsection{\textbf{First Execution}}
\label{subsubsec:trace-first}
In this execution run, \texttt{DynaCI} attaches four probes to the developer's local environment to partition commands and trace their durations, file accesses, and package dependencies with \texttt{strace}~\cite{kranenburg_1991_strace}, which we detail below. 
\begin{itemize}[leftmargin=1em]
    \item \textbf{Command probe.} Developers typically execute a sequence of commands to prepare, build, test, and validate a project in their local environment. While these commands collectively accomplish multiple tasks, their boundaries are not explicitly demarcated. This poses a challenge when synthesizing CI workflows, where commands must be organized into semantically meaningful workflow steps~\cite{ghadocumentation,Mazrae2026}. To address this problem, \texttt{DynaCI} introduces a Command probe that analyzes developer command sequences and partitions them into executable units that can later be mapped to workflow steps. Each unit is a command or expression node that is defined at the top-level of a command sequence's parse tree~\cite{dong_2023_bash}.
    \item \textbf{Duration probe.} We develop a Duration probe that captures the durations of commands identified by a Command probe, i.e., the datetime the execution completes minus the datetime the execution starts.
    \item \textbf{File probe.} We develop a File probe that captures references to files in the commands' system calls by tracing system calls that use file paths, such as \texttt{open} and \texttt{openat}~\cite{manual_2025_open}.
    \item \textbf{Package probe.} If a workflow runs developer commands, it must install their packages, or dependencies. We develop a Package probe that identifies these packages by checking if any of their associated file paths have been accessed.
\end{itemize}

\begin{table}[t]
    \centering
    \small
    \caption{Example Runtime Traces Extracted by the Probes During the First Execution}
    \label{tab:first-run}
    \resizebox{\columnwidth}{!}{\begin{tabular}{p{2cm} p{2.5cm} p{7cm}}
    \toprule
    \textbf{Probe} & \textbf{Captured Field} & \textbf{Example Extracted Data} \\
    \midrule
    
    \multirow{2}{*}{Command Probe} 
    & Command& make ci\\
 & LoC&1\\
    \midrule
    \multirow{3}{*}{Duration Probe} & Start Time& 2025-07-20 23:19:46\\
 & End Time&2025-07-20 23:21:10\\
 & Duration&84 seconds\\
    
    \midrule
    
    File Probe 
    & Accessed Files 
    & /app/tests/test\_requests.py\\
    
    \midrule
    
    \multirow{2}{*}{Package Probe} 
    & apt Packages& python3==3.10.12\\
    & pip Packages&pytest==2.8.0\\
    \bottomrule
    \end{tabular}}
\end{table}

By the end of the first execution, the four probes have extracted runtime traces, describing the behavior of each executed command. \autoref{tab:first-run} shows an example of the information extracted for the \texttt{psf/requests} project.

\subsubsection{\textbf{Second Execution}}
\label{subsubsec:trace-second}
In this execution, \texttt{DynaCI} traces access to environment variables, since they can have substantial impacts on the behavior of software systems~\cite{rahman_2025_env}.
\texttt{DynaCI} uses the following probes implemented for the shell and Python interpreter:
\begin{itemize}[leftmargin=1em]
    \item \textbf{Shell environ probe.} This probe captures environment variable (i.e., environ\footnote{environ is a global variable provided by the C standard library that holds the environment variables on Linux~\cite{python_2025_environ}.}) accesses by system library calls, such as \texttt{getenv}~\cite{manual_2025_getenv}, with \texttt{ltrace}~\cite{cespedes_2019_ltrace}.
    \item \textbf{Python environ probe.} This probe captures environment variable accesses by a language-specific runtime interface. Many runtimes, such as the Python interpreter and the Java Virtual Machine, provide an abstraction over the underlying OS environment. The probe instruments this abstraction to record environment variables that are accessed during execution. For Python, we implement the probe using a site-specific customization script that leverages the \texttt{environ} interface in the \texttt{os} module to capture environment variables during interpreter initialization~\cite{python_2025_sitecustomize}. This approach can be readily extended to other languages; for example, in Java, a similar probe can be implemented using agent-based instrumentation via \texttt{java.lang.instrument}~\cite{peldszus_2024_java}.
\end{itemize}

\subsection{Apply Heuristics}
\label{subsec:approach-build}
In this phase, \texttt{DynaCI} applies heuristics to runtime information collected from the developer's local environment (e.g., commands, dependencies, and environment variables) to construct an initial workflow that will be refined later.

\subsubsection{\textbf{Filter Candidate Workflow Components}}
\label{subsubsec:build-component} 
As of December 2025, GHA supports 84 workflow components, each representing a configurable element of a workflow~\cite{github_2025_syntax}.
However, some components are CI-specific and do not correspond to configurations that exist in a developer’s local environment. 
Since \texttt{DynaCI} aims to generate workflows from local environments, it avoids using components that do not have local counterparts unless their inclusion is required to run the workflow. 
We grouped the 84 workflow components into eight categories using official GHA documentation~\cite{github_2025_syntax}.
For three of the most important categories, below, we discuss our inclusion and exclusion criteria for identifying candidate workflow components.

\textbf{Environment (27/84)}, i.e., the components that define key-value pairs that can be accessed by other workflow components: These pairs are either provided to the workflow as secrets, i.e., confidential settings that are not available to our research or the general public (21), or as hard-coded values in the workflow's file (6). 
The latter typically specify environment variables required by the workflow, a job, or an individual step. Since GitHub-provided inputs are not available during local execution, \textit{\texttt{DynaCI} can only add environment variables that are observable in a local execution environment.}

\textbf{Execution (10/84)}, i.e., the components that interact with the OS or shell context: These components either configure the OS and shell environment~(7) or execute commands within these environments (3). 
We retain the command-execution components because they directly correspond to commands that developers run in their local environments. 
We exclude the former components from \texttt{DynaCI} workflows since their usage in real-world workflows is exceedingly rare. 
For example, the former were only 0.48\% of our \textit{CI Workflow Dataset} (\autoref{subsubsec:build-parametric}). \textit{\texttt{DynaCI}'s workflows thus only include Execution components that directly execute commands.} 

\textbf{Worker (8/84)}, i.e., the components that manage when jobs and steps are started or stopped: These components, order job executions (4), start jobs that meet a condition (2), and timeout job or step executions (2). Because conditions that trigger jobs are CI-specific, \textit{\texttt{DynaCI}'s workflows can only include Worker components that order and timeout job and step executions.}

\subsubsection{\textbf{Apply Non-Parametric Heuristics}}
\label{subsubsec:build-nonparametric}
Using the runtime information extracted in Section~\ref{subsec:approach-trace}, \texttt{DynaCI} identifies the candidate components discussed in Section~\ref{subsubsec:build-component}. It then applies the following \textit{non-parametric heuristics}---rule-based decisions derived directly from the observed runtime behavior---to generate a preliminary workflow file. These heuristics rely solely on the runtime information collected from the developer’s environment and do not depend on tunable parameters.\footnote{Note that some heuristics require parameters whose values must be derived empirically; these \textit{parametric heuristics} are described in Section~\ref{subsubsec:build-parametric}.}

\textbf{For each job, set the environment variables extracted from the runtime traces:}
In GHA, \textbf{\texttt{jobs.$<$id$>$.env}} defines environment variables accessible to all steps in a job. Based on runtime information, \texttt{DynaCI} initializes this component with variables accessed in the developer’s local environment.

\textbf{For each job, set the configuration matrix using the programming language version and OS name extracted from runtime traces:}
In GHA, \textbf{\texttt{jobs.$<$id$>$.strategy.matrix}} defines a matrix of configurations under which a job runs. Most matrices cover a range of programming language versions, such as Python 3.10, 3.11, and 3.12, and specify the machines on which the workflow runs, such as GitHub-hosted Ubuntu, Windows, and MacOS runners. \texttt{DynaCI} defines the matrix in each job using the OS and Python version observed in runtime traces. However, \texttt{DynaCI} currently relies on Linux-specific tools (e.g., \texttt{strace}, \texttt{ltrace}) to capture runtime information, so it only generates workflows for the GHA Ubuntu runner, accounting for 64.6\% of workflow jobs in GHA in general~\cite{bouzenia_2024_gha}.

\textbf{For each job, set the execution machine using the OS defined in the configuration matrix:}
In GHA, the machine on which a job runs is specified by \textbf{\texttt{jobs.$<$id$>$.runs-on}} component. \texttt{DynaCI} configures the value of this component with the OS version mentioned in \texttt{jobs.$<$id$>$.strategy.matrix}.

\textbf{For each job, derive workflow steps from runtime information:}
In GHA, the sequence of tasks that a job performs is specified by \textbf{\texttt{jobs.$<$id$>$.steps[*]}}. According to runtime information, \texttt{DynaCI} adds a step to each job that installs the \texttt{apt} and \texttt{pip} packages that are missing in the developer's local dependency configuration file (e.g., \texttt{requirements.txt}). \texttt{DynaCI} also groups the developer's commands into separate steps. 
First, commands (from the Command probe) are partitioned into candidate step groups, preserving order and ensuring each step contains at least one command.
Second, \texttt{DynaCI} determines the total size, in LoC, and duration of each group using the runtime information extracted from the Command probe and the Duration probe. 
Third, it uses Multi-Criteria Decision Making (MCDM)~\cite{pymoo_2004, veljic_2026_mcdm} to find the step group that minimizes the spread of command durations across its steps while also maximizing the total number of steps. This is based on our intuition that there are \textit{primary} commands, which perform the main activities and thus run longer (e.g., \texttt{pytest}), that are surrounded by shorter \textit{secondary} commands, which provide support (e.g., \texttt{echo}).

\textbf{For each job, include reusable actions to prepare the workflow environment:}
In GHA, actions or reusable tasks are invoked as steps using the \textbf{\texttt{jobs.$<$id$>$.steps[*].uses}} component.
\texttt{DynaCI} always adds \texttt{actions/checkout}~\cite{github_2018_checkout} so that the workflow can access its repository. If the workflow uses Python or the \texttt{uv} package manager, \texttt{DynaCI} will also add \texttt{actions/setup-python}~\cite{github_2018_setup-python} or \texttt{astral-sh/setup-uv}~\cite{github_2024_setup-uv}, respectively. These actions set up program dependencies in the workflow environment.

\subsubsection{\textbf{Apply Parametric Heuristics}}
\label{subsubsec:build-parametric}
Some of GHA components require parameter information that cannot be extracted from runtime traces, and are CI- or project-specific (e.g., setting timeout and concurrency parameters). 
In our study, we classify such parameters into two categories: threshold parameters and multiplier parameters. 
A \textit{threshold parameter} specifies the minimum execution duration a job must have before a workflow component is included (e.g., enabling concurrency and timeout settings). 
This allows \texttt{DynaCI} to add such GHA components into a workflow only for potential longer-running jobs, where such components are more useful.
A \textit{multiplier parameter} determines the numeric value of a workflow component by scaling the observed job duration (e.g., setting timeout limits once the timeout component is included). 
This allows \texttt{DynaCI} to determine the numerical value for setting certain components that are related to a job's runtime duration, i.e., the sum of the durations of corresponding commands executed in the developer's local environment, as reported by the Duration probe. 

We derived these parameters from Cardoen et al.'s dataset of 2,595,399 GHA workflows~\cite{cardoen_2024_gha}. We first filtered this dataset by removing invalid workflows, old versions of valid workflows, and workflows that are unrelated to Python-based CI. We consider a workflow to be valid if it conforms to a GHA workflow schema~\cite{workflow_schema}, and a workflow is Python-based if its repository's most popular language is Python according to the GitHub REST API~\cite{github_2022_languages}. Doing so resulted in 15,285 workflows, which we call the \textit{CI Workflow Dataset}. Setting aside 50 workflows with the highest star counts for our \textit{Evaluation Dataset} (\autoref{subsec:evaluation-setup}), we used the GitHub REST API to extract the last successful job runs, if any, from the remaining 15,235 workflows. This left us with 7,157 job runs, which we refer to as the \textit{Heuristic-Tuning Dataset}. 
For the workflow components below, we extracted the durations of job runs that use them, removed their outliers, and treated their averages as default threshold and multiplier values. 
Below, we list these parameter values and their associated heuristics, which \texttt{DynaCI} uses to construct workflows. 

\textbf{Set a timeout limit for job execution (Threshold = 319s, Multiplier = 13):} In GHA, \textbf{\texttt{jobs.$<$id$>$.timeout-minutes}} specifies the maximum number of minutes that a job can run before it is forced to stop. It is included in the workflow if the corresponding job duration observed in the developer’s local environment exceeds 319 seconds; otherwise, it is excluded. When included, its value is set to 13 times the corresponding job duration observed in the developer’s local environment. 

\textbf{Set concurrency control to prevent long-running jobs from executing simultaneously (Threshold = 138s, Multiplier = N/A):}
In GHA, \textbf{\texttt{jobs.$<$id$>$.concurrency}} prevents jobs that are in the same concurrency group from running at the same time. These groups are defined as strings and often include context variables, or useful values that GHA substitutes into workflows before running them~\cite{github_contexts}. If included, \texttt{DynaCI}, using contexts, sets its concurrency group to the name of the workflow followed by the ref that triggered it (i.e., \$\{\{ github.workflow \}\}-\$\{\{ github.ref \}\}). This prevents multiple, longer-running jobs from starting at once.

\textbf{Set fail-fast behavior to allow all jobs in a matrix to run even if one fails (Threshold = 66s, Multiplier = N/A):}
In GHA, \textbf{\texttt{jobs.$<$id$>$.strategy.fail-fast}}, if set to true, will cancel all jobs in a configuration matrix if any job in that matrix fails. It is included in the workflow and set to "false" if its job's duration exceeds the threshold; otherwise, it is not included, and the value would be "true" by default.

By the end of this phase, \texttt{DynaCI} is expected to result in \textit{\textbf{an initial workflow}} file that can then be refined via LLMs.

\subsection{Refine with LLMs}
\label{subsec:approach-refine}
To ensure that there are no missing components, \textit{Apply Heuristics} step is designed to add components as much as possible to the workflow. However, this over-permissive design can make the initial workflow generated by \texttt{DynaCI} overly complicated and error-prone. To mitigate this issue, \textit{Refine with LLMs} step is designed to be more restrictive. It uses an LLM to analyze runtime information, remove unnecessary and/or problematic workflow components, and patch any syntax errors that arise during refinement. 

To support this process, we extract runtime information about the GHA runner that the workflow is expected to run in, providing it to \texttt{DynaCI} as CI context (\autoref{subsubsec:extract-extra-info}).
For a given initial workflow, \texttt{DynaCI} separates all the available runtime information into chunks and encodes the chunks into embeddings. (\autoref{subsubsec:refine-embed}). Then, \texttt{DynaCI} retrieves the most relevant chunks via Retrieval-Augmented Generation (RAG)~\cite{fan_2024_rag} to provide the LLM runtime context. \texttt{DynaCI} finally refines the content of the workflow with the LLM on two levels: the job-level (\autoref{subsubsec:refine-job}) and workflow-level (\autoref{subsubsec:refine-workflow}).

\subsubsection{\textbf{Extract Runtime Environment from GitHub Actions}}
\label{subsubsec:extract-extra-info} 
In every initial workflow, \texttt{DynaCI} sets environment variables and installs packages found in the developer's local environment. However, in many cases, these features may already be found on a GHA runner. To minimize redundancies and potential conflicts, \texttt{DynaCI} provides runtime information about the target runner environment to the LLM as additional context for refinement. We acquired this information manually by running \texttt{apt}, \texttt{pip}, and \texttt{env} commands on the target runner.

\subsubsection{\textbf{Embed Runtime Information}}
\label{subsubsec:refine-embed}

The runtime information available from local and CI environments can exceed $1$ GB in size, so \texttt{DynaCI} cannot pass it all to the LLM without exploding costs and degrading performance~\cite{xu_2024_llm, du_2025_llm}.  Instead, \texttt{DynaCI} chunks and embeds the information in a database, only sending the most relevant chunks to the LLM (i.e., RAG).

\subsubsection{\textbf{Refine at Job-Level}}
\label{subsubsec:refine-job}
Because each job reflects a different entry point in the developer's local environment, \texttt{DynaCI} traces them separately. Thus, each job has its own runtime information. 
Instead of refining the entire workflow at once, \texttt{DynaCI} uses a \textbf{\textit{divide-and-conquer}} approach; it extracts the jobs from the initial workflow and improves them separately (\textbf{\textit{divide}}).
For each job, \texttt{DynaCI} first retrieves runtime chunks with cosine similarity $\geq 60\%$ of its run blocks.
Next, \texttt{DynaCI} sends a job prompt with these chunks to an LLM with instructions to remove unneeded components from the job (e.g., environment variables, package installations) and to substitute direct references to OS and Python versions with the job's matrix variables, which is a best practice~\cite{github_2025_workflows}. 
The prompt includes information about the GHA environment that the job will run in (\autoref{subsubsec:extract-extra-info}) and warnings against making additional changes, as preliminary experiments showed a tendency for the LLM to hallucinate beyond the intended scope. 

\subsubsection{\textbf{Merge Refined Jobs and Refine at Workflow-Level}}
\label{subsubsec:refine-workflow}
After \texttt{DynaCI} receives the refined job blocks, it instructs the LLM to merge them into a single workflow (\textbf{\textit{conquer}}). 
To mitigate hallucinations, \texttt{DynaCI} validates the generated workflow using \texttt{actionlint}, a static analyzer for GHA workflows~\cite{github_2021_actionlint}. 
If \texttt{actionlint} returns an error, \texttt{DynaCI} adds the error message to the workflow prompt and regenerates the refined workflow. It repeats this process until \texttt{actionlint} no longer produces an error, or five refinement iterations have occurred, producing the \textbf{\textit{final workflow}}. 

%% file: 4-Evaluation.tex
\section{Experimental Setup}
\label{subsec:evaluation-setup}
To evaluate the performance of \texttt{DynaCI}, we required a dataset that establishes a correspondence between CI workflows and their counterparts in local environments. We began with our \textit{Evaluation Dataset} of 50 workflows with the highest star counts that we keep aside during \autoref{subsubsec:build-parametric} to ensure that there is no data leakage between our \textit{Heuristic-Tuning Dataset} and the evaluation setup. For each project in our sample, we then consulted the projects' source code and documentation to create representative local environments (\autoref{subsubsec:evaluation-environment}). Within each environment, we generate a workflow file using \texttt{DynaCI} (\autoref{subsubsec:evaluation-dynaci}) as well as the state-of-the-art workflow generation tools, \texttt{GH-WCOM}~\cite{mastropaolo_2024_synthesis} and \texttt{LLM4GW}~\cite{ghaleb_2025_synthesis} (\autoref{subsubsec:evaluation-baselines}). We finally compare these workflows against their targets on effectiveness, efficiency, and their need for modification to answer RQ1--RQ3 (\autoref{subsec:evaluation-metrics}).

\subsection{\textbf{Local Environment Configuration}}
\label{subsubsec:evaluation-environment}
Two authors manually inspected each projects' developer-facing documentation (e.g., \texttt{CONTRIBUTING.md}) and configurations (e.g., \texttt{pyproject.toml}) to identify the commands that developers use to set up their local environments. For example, in \texttt{httpie/cli}, the \texttt{CONTRIBUTING.md} file mentions test-related commands, such as \texttt{make test} and \texttt{python -m pytest}. We found 90.8\% of commands in our sample involve similarly generic commands, such as \texttt{tox}. To ensure that the reconstructed local environments were executable and representative of developer machines, we also added setup commands that were not always explicitly documented in the source code or project documentation. 
For instance, when needed, we installed project dependencies using commands such as \texttt{pip install -r requirements.txt} before executing the documented test commands. 
Previous works used similar approaches to identify build and test commands~\cite{bouzenia_2025_executionagent} as well as dependency installation commands~\cite{yu_2025_cxxcrafter} in documentation and configuration files.

To ensure the reproducibility of the aforementioned environments, we defined each environment in a Dockerfile that performs the following key steps. First, the Dockerfile clones the corresponding repository into the container. Second, it installs \texttt{DynaCI} and its dependencies (i.e., \texttt{strace}, \texttt{ltrace}, and \texttt{actionlint}). Third, the Dockerfile installs the most recent Python version that the project supports as well as the package  manager that the project uses, if any (e.g., \texttt{uv}). Fourth, it sets any environment variables with values mentioned in the project's source code and documentation. 
Lastly, the Dockerfile executes developer commands that we identified with the aforementioned approach, and which a developer would provide as an input file to \texttt{DynaCI} (Section~\ref{subsec:approach-trace}).

\subsection{\textbf{Workflow File Generation Using \texttt{DynaCI}}}
\label{subsubsec:evaluation-dynaci}
Within each Dockerized environment, \texttt{DynaCI} is configured to execute each of the developer's commands. 
Before their execution, \texttt{DynaCI} attaches its probes into environment to capture its runtime information. 
\texttt{DynaCI} transforms this runtime information into an \textbf{\textit{initial workflow}} file, which then would be refined via an LLM to create the \textbf{\textit{final workflow}} file. 
We treat this final workflow file as the output of \texttt{DynaCI} when comparing it with the baseline approaches.

\subsection{\textbf{Workflow Generation Using Baseline Approaches}}
\label{subsubsec:evaluation-baselines}
To contextualize \texttt{DynaCI}'s performance, we generate workflows using several (state-of-the-art) approaches:

\textbf{Commands-Only, i.e., using only the developer's commands to generate a workflow:} For each project, we instruct a LLM to produce a GHA workflow \textit{only} using the commands that developers would be run in their Dockerized environment. This baseline evaluates whether commands alone are enough to produce an acceptable workflow with an LLM.

\textbf{Heuristics-Only, i.e., using \texttt{DynaCI}'s heuristics to generate an initial workflow without refinement:} Specifically, it follows \texttt{DynaCI}'s \textit{Trace Commands} step (\autoref{subsec:approach-trace}) and \textit{Apply Heuristics} step (\autoref{subsec:approach-build}) to produce an initial workflow, omitting the subsequent LLM-based refinement. This baseline results in more permissive workflows that can incorporate improbable components, which the refinement step might have otherwise removed. We include this approach in our paper as \textbf{\textit{an ablation study}} to check whether the refinement step in \texttt{DynaCI} makes measurable improvements.

\textbf{\texttt{GH-WCOM}~\cite{mastropaolo_2024_synthesis}, i.e., a transformer-based approach that provides suggestions for auto-completing GHA workflows:} \texttt{GH-WCOM} is trained on two tasks: (i) \textit{Next Statement (NS)}, which predicts the next step in a job given the preceding workflow context, and (ii) \textit{Job Completion (JC)}, which generates the remaining steps of a job given their names. 
Since \texttt{GH-WCOM} requires incomplete workflows, we remove runtime-based components, such as steps, from workflows generated by \texttt{Heuristics-Only}.
We then treat workflow generation as an NS task, sequentially predicting each job’s next steps until the original number of steps is restored.

\textbf{\texttt{LLM4GW}~\cite{zhang_2024_synthesis}, i.e., a set of prompt optimization techniques for tasks related to GHA:} \texttt{LLM4GW} introduces five prompt templates (P1--P5) for workflow generation tasks, which include an increasing amount of information about the workflow, jobs, and steps to be generated. 
However, our use case assumes that developers cannot articulate details about their target workflows, so only prompt P1 of \texttt{LLM4GW}, which describes workflow-level information and job IDs, is comparable to \texttt{DynaCI}. 
We use their tool to convert the input in \texttt{GH-WCOM}~\cite{mastropaolo_2024_synthesis} into a P1 prompt, providing it to the same LLM that \texttt{DynaCI} uses to generate a workflow.

\subsection{Performance Metric Computation}
\label{subsec:evaluation-metrics}

\subsubsection{\textbf{RQ1. Effectiveness}}
\label{subsubsec:evaluation-effectiveness_metrics}
We used textual similarity metrics to evaluate how effectively \texttt{DynaCI} performs against its baselines. For each tool, we computed the average of the following scores across our 50 projects: 
\begin{itemize}[leftmargin=1em]
    \item \textbf{BLEU Score} compares two texts on a token-level. It is commonly used to evaluate program synthesis tools in general~\cite{wang2021codet5, Li2023ICSE, liu2024reliability, xiong2024program}, and CI workflow generation tools in particular~\cite{mastropaolo_2024_synthesis, zhang_2024_synthesis}.
    \item \textbf{F1 Score} captures how well a given tool balances correctness and completeness in including a component in a workflow. Key-value matching metrics, such as the F1 score, are commonly used to evaluate YAML generation tasks~\cite{xu_2024_metric, zhang_2024_synthesis, santolucito_2018_metric}. To compute this, we first record the number of true-positive (TP), false-positive (FP), and false-negative (FN) components in each generated workflow. For example, a component is TP if it is found in both the target workflow (i.e., developer-written workflow) and the generated workflow and the values match. We then use these values to calculate the precision and recall, of which F1 score is the harmonic mean.
\end{itemize}

\subsubsection{\textbf{RQ2. Efficiency}}
\label{subsubsec:evaluation-cost_metrics}
We recorded the time taken by each tool to generate each workflow. If the tool used an LLM, we also record the number of tokens it consumed. Together, these metrics characterize the efficiency of \texttt{DynaCI} and its baselines.

\subsubsection{\textbf{RQ3. Modification}}
\label{subsubsec:evaluation-error_metrics}
Similarity between generated and target workflows (RQ1) does not guarantee similar execution. 
To measure the latter, two authors (with two years of CI research experience as of December 2025) manually executed all generated workflows in forked projects, iteratively revising them until their execution logs~\cite{github_logs} matched the target workflows.
The authors recorded each LoC changed in generated workflows and, from these revisions, derived common repair categories. 
We used this information to calculate the following metrics for \texttt{DynaCI} and its baselines: 
\begin{itemize}[leftmargin=1em]
    \item \textbf{Repair Rate} measures the proportion of tool-generated workflows that are missing functionality (e.g., builds, test cases, etc.) in a generated workflow and thus require repairs.
    \item \textbf{Delta LoC} refers to the total LoC inserted, modified, or deleted to implement missing functionality in a workflow.
\end{itemize}

\subsection{Hardware / Software Requirements}
The Dockerized environments for \texttt{DynaCI} and \texttt{Heuristic-Only} were hosted on a Ubuntu 24.04 machine with an Intel i5-10600K CPU @ 4.10GHz CPU, a Radeon RX 5700 XT GPU, and 16 GB RAM. We used OpenAI's GPT-4o-mini~\cite{openai_2024_gpt-4o-mini} for generation and ADA-002~\cite{openai_2022_ada-002} for embeddings, though neither model is intrinsic to our approach and can be replaced with any suitable model.

\section{Results}
\label{sec:results}

\subsection{RQ1. How effective is \texttt{DynaCI} at generating CI workflows?}
\label{subsec:evaluation-effectiveness}

\begin{table}[b]
    \caption{Average Performance of Workflow Generation Tools}
    \label{tab:rq1-scores}
    \centering
    \resizebox{0.7\columnwidth}{!}{
    \begin{tabular}{l|l|llll}
        \toprule
        \textbf{Tool} & \textbf{BLEU} & \textbf{F1} & \textbf{TP} & \textbf{FP} & \textbf{FN} \\
        \midrule
        \texttt{DynaCI} & 52.81 & 62.06 & 14.22 & 5.00 & 13.50 \\
        \texttt{Heuristics-Only} & 21.31 & 48.17 & 14.70 & 24.64 & 13.02 \\
        \texttt{GH-WCOM}~\cite{mastropaolo_2024_synthesis} & 31.68 & 41.10 & 8.04 & 5.88 & 19.68 \\
        \texttt{LLM4GW}~\cite{zhang_2024_synthesis} & 11.99 & 18.65 & 3.00 & 6.34 & 24.72 \\ 
        \texttt{Commands-Only} & 13.76& 36.72& 7.88& 5.12& 19.84\\
    \bottomrule
    \end{tabular}}
\end{table}

\begin{table}[tb]
    \centering
    \caption{Breakdown of the Average F1 Scores}
    \resizebox{\columnwidth}{!}{
    \begin{tabular}{ll|ccccc}
        \toprule
        \textbf{Component} & \textbf{Type} &  \textbf{\texttt{DynaCI}}  &\textbf{\texttt{Heuristics-Only}} & \textbf{\texttt{GH-WCOM}} & \textbf{\texttt{LLM4GW}}  &\textbf{\texttt{Commands-Only}} \\
        \midrule
        jobs.$<$id$>$.env.* & Environment &  2.58  & 3.73 & 0.00 & 0.00 & 0.00\\
        jobs.$<$id$>$.needs & Worker &  0.00  &0.00 & 0.00 & 20.00 & 0.00\\
        jobs.$<$id$>$.runs-on & Execution &  96.05  &96.05 & 96.05 & 1.95 & 1.67\\
        jobs.$<$id$>$.steps[*].run & Execution &  72.36  &67.32 & 7.41 & 9.61 & 63.57\\
        jobs.$<$id$>$.steps[*].uses & Miscellaneous &  96.80  &96.80 & 91.70 & 90.51 & 66.02\\
        jobs.$<$id$>$.steps[*].with.* & Environment &  71.65  &71.65 & 71.65 & 1.00 & 0.00\\
        jobs.$<$id$>$.strategy.fail-fast & Error & 32.89 &32.89 & 0.00 & 0.00 & 0.00\\
        jobs.$<$id$>$.strategy.matrix.* & Environment &  45.97  &47.94 & 47.94 & 0.29 & 0.00\\
        \bottomrule
    \end{tabular}}
    \label{tab:rq1-breakdown}
    \vspace{-1em}
\end{table}

\autoref{tab:rq1-scores} shows the textual similarity metrics for \texttt{DynaCI} and our baselines on our evaluation set.

\textbf{Observation 1: \texttt{DynaCI} demonstrates improved performance over the baselines across BLEU and F1-scores.}
We observe that \texttt{DynaCI} outperforms the four baseline approaches in BLEU, achieving a +21.13\% higher score on average. We also found that \texttt{DynaCI} has the highest average F1 score of 62.06\%, outperforming all four baselines by +13.89\% and just the state-of-the-art by +20.96\%.

The table reports TPs, FPs, and FNs to show how \texttt{DynaCI}'s refinement step improves over \texttt{Heuristics-Only}. The \texttt{Heuristics-Only} approach, which corresponds to \texttt{DynaCI}'s \textit{Apply Heuristics} step, is more permissive: it includes as many runtime-observed components as possible, but also introduces extraneous components, increasing FPs.
For example, in \texttt{yt-dlp/yt-dlp}, \texttt{Heuristics-Only} added an environment variable, NO\_PROXY, from  runtime information to a workflow. 
However, this variable prevented network traffic from routing through GHA proxy servers, resulting in an error. 
\texttt{DynaCI} mitigates this issue with its more restrictive \textit{Refine with LLMs} step, which excludes as many extraneous components, such as NO\_PROXY, from the \textit{Apply Heuristics} step as it can.
While this added step decreases FPs, it risks removing important components in the process, decreasing TPs and increasing FNs. 
However, \texttt{DynaCI} still manages to balance TPs, FPs, and FNs with these two steps, improving F1-scores overall.  
We used a two-sample t-test to compare F1 scores between \texttt{DynaCI} and the baselines; all differences in \autoref{tab:rq1-scores} are statistically significant ($p < 0.05$).

\autoref{tab:rq1-breakdown} reports average component-level F1 scores for components where tools differ. Below are the main observations.

\textbf{Observation 2: \texttt{DynaCI} and \texttt{Heuristics-Only} generally perform better on components related to environment variables, error handling, and command execution.}
For example, the \textbf{jobs.$<$id$>$.steps[*].run} component runs the developer's commands within a shell environment. \texttt{GH-WCOM}, \texttt{LLM4GW}, and \texttt{Commands-Only} often exclude these developer's commands (\autoref{subsec:evaluation-error}) and thus have poor F1 scores. \texttt{Heuristics-Only} adds \texttt{apt} and \texttt{pip} packages that the developer's local environment accesses. \texttt{DynaCI} does the same but removes packages that are already installed in the GHA runner environment with an LLM. 
Since these package-installation statements are absent from the target workflows, they are counted as FPs, reducing both tools' F1 scores.

Next, \texttt{DynaCI} and \texttt{Heuristics-Only} approach outperforms the others with respect to the \textbf{jobs.$<$id$>$.strategy.fail-fast} components, which stops running a job's matrix if a single job is unsuccessful. While the other tools do not include this component, \texttt{DynaCI} and \texttt{Heuristics-Only} do if the duration of the developer's commands exceeds a certain threshold (\autoref{subsubsec:build-parametric}).
We also find that the \textbf{jobs.$<$id$>$.env.*} component, which defines environmental variables across steps, exhibits low overall performance. However, \texttt{DynaCI} and \texttt{Heuristics-Only} outperform the other tools by defining all environmental variables from their runtime information, including many that are FPs. Since \texttt{GH-WCOM}, \texttt{LLM4GW}, and \texttt{Commands-Only} do not define any environment variables, their F1 scores are 0. 

\texttt{DynaCI},  \texttt{Heuristic-Only}, and \texttt{GH-WCOM} consistently outperform \texttt{LLM4GW}~\cite{zhang_2024_ci} and \texttt{Commands-Only}, with respect to components that define execution configurations, such as \textbf{jobs.$<$id$>$.runs-on}, \textbf{jobs.$<$id$>$.strategy.matrix.}, and \textbf{jobs.$<$id$>$. steps[*].with.*}.
This is because, unlike other tools, \texttt{LLM4GW} and \texttt{Commands-Only} does not define or use \texttt{strategy.matrix} to parameterize configurations. 
As a result, it fails to generate multiple configuration variants or consistently propagate values to dependent components, such as \texttt{runs-on} and \texttt{steps[*].with}. 
Since \texttt{LLM4GW} and \texttt{Commands-Only} rely on hard-coded job configurations, they produce missing variants, inconsistencies, and FNs.

\subsection{RQ2. How time and cost efficient is \texttt{DynaCI}?}
\label{subsec:evaluation-efficiency}
\begin{table}[tb]
    \centering\caption{Median Duration of Tools (in Seconds)}
    \resizebox{0.85\columnwidth}{!}{
    \begin{tabular}{l|c|cc|c|cc}
        \toprule
        \multicolumn{2}{c|}{} & \multicolumn{2}{c|}{\textbf{Trace Commands}} & \textbf{Apply Heuristics} & \multicolumn{2}{|c}{\textbf{Refine with LLMs}} \\
        \midrule
        \textbf{Tool} & \textbf{Total} & Trace & Parse & Generate & Embed & Refine \\
        \midrule
        \texttt{DynaCI} & 333.66 & 147.92 & 39.52 & 0.01 & 133.01 & 13.20 \\
        \texttt{Heuristics-Only} & 187.45 & 147.92 & 39.52 & 0.01 & - & - \\
        \texttt{GH-WCOM} & \~0 & - & - & \~0 & - & - \\
        \texttt{LLM4GW} & \~0 & - & - & \~0 & - & - \\
        \texttt{Commands-Only}& \~0 & - & - & \~0 & - &- \\
        \bottomrule
    \end{tabular}}
    \label{tab:rq2-durations}
    \vspace{-0.5em}
\end{table}

\textbf{Observation 3: \texttt{DynaCI} synthesizes a workflow within 5 minutes and 33 seconds on average.} 
\autoref{tab:rq2-durations} reports the average workflow-generation time for \texttt{DynaCI} and our baselines on our evaluation set. 
\texttt{Commands-Only}, \texttt{GH-WCOM}, and \texttt{LLM4GW} had negligible durations. \texttt{Heuristics-Only} and \texttt{DynaCI} took longer, 187.45 and 333.66 seconds, respectively, due to their additional steps. The longest step is \textit{Trace Commands}, which monitor the developer's local environment twice. 
Since designing a workflow from scratch typically takes much longer than five minutes and 33 seconds for a developer~\cite{zhang_2024_ci, valenzuela-toledo_2022_gha, saroar_2023_gha}, all tools, including \texttt{DynaCI}, are time-efficient.

\begin{table}[t]
    \centering
    \caption{Median Token Cost per Workflow of Tools}
    \vspace{-0.1cm}
    \resizebox{0.5\columnwidth}{!}{
    \begin{tabular}{l|c|cc}
        \toprule
        \textbf{Tool} & Total & Embed & Prompt \\
        \midrule
        \texttt{DynaCI} & 653,755& 568,734& 85,021\\
        \texttt{Heuristics-Only} & - & - & - \\
        \texttt{GH-WCOM} & - & - & - \\
        \texttt{LLM4GW} & 199 & - & 199 \\
        \texttt{Commands-Only}& 868& - &868\\
        \bottomrule       
    \end{tabular}}
    \label{tab:rq2-cost}
    \vspace{-1em}
\end{table}
\textbf{Observation 4: It generally costs 653,755 tokens to synthesize a workflow with \texttt{DynaCI}.} 
\autoref{tab:rq2-cost} reports the average token cost per workflow. 
\texttt{Commands-Only} and \texttt{LLM4GW} consume only 868 and 199 tokens per workflow, respectively, compared to \texttt{DynaCI}'s 653,755 tokens.  
This is because \texttt{DynaCI} embeds and prompts using runtime information chunks, which account for most of its token usage.
However, this cost is modest because most repositories use only a few CI workflows~\cite{decan_2022_gha}, and the performance gains of \texttt{DynaCI} outweigh this trade-off. 
\texttt{GH-WCOM} and \texttt{Heuristics-Only} do not use LLMs and therefore incur no token cost.

\subsection{RQ3. How much modification is needed for CI workflows generated by \texttt{DynaCI}?}
\label{subsec:evaluation-error}
\begin{table}[t]
    \centering
    \caption{Median / Total Delta LoC by Change Type}
    \vspace{-0.1cm}
    \resizebox{0.65\columnwidth}{!}{
    \begin{tabular}{l|c|ccc}
        \toprule
        \textbf{Tools} & \textbf{Total Changes} & \textbf{Insert} & \textbf{Delete} & \textbf{Modify} \\
        \midrule
        \texttt{DynaCI} & 4 / 52 & 2 / 28 & 1 / 11 & 1 / 13 \\
        \texttt{Heuristics-Only} & 4 / 83 & 2 / 23 & 1 / 16 & 1 / 44 \\
        \texttt{GH-WCOM}  & 10 / 592 & 5 / 340 & 4 / 200 & 1 / 52 \\
        \texttt{LLM4GW} & 10 / 592 & 5 / 340 & 4 / 200 & 1 / 52 \\
        \texttt{Commands-Only}& 5 / 303 & 3 / 221 & 1 / 36 & 1 / 46 \\
        \bottomrule
    \end{tabular}}
    \label{tab:rq3-loc}
    \vspace{-0.5em}
\end{table}

For the \texttt{DynaCI}-generated workflows that are missing functionality from their target, \autoref{tab:rq3-loc} shows the median and total Delta LoC between the generated and target workflows. 

\textbf{Observation 5: \texttt{DynaCI}'s \textit{Refine with LLMs} step removes many problematic components that its \textit{Apply Heuristics} step introduces}.
\texttt{DynaCI} requires the fewest total changes (52), while having the same median number of changes (4) as \texttt{Heuristics-Only}.
This difference in total changes is due to the absence of the \textit{Refine with LLMs} step in the \texttt{Heuristics-Only} approach. 
For example, \texttt{Heuristics-Only} attempts to install all \texttt{apt} packages in the developer's local environment. However, there were many cases in which a package, such as a specific Python version, could not be found due to a missing package repository in the GHA environment. 
These unsupported installations often need to be removed manually, whereas \texttt{DynaCI} removes many of them during LLM refinement.

\textbf{Observation 6: Developers must change substantially fewer LoC to repair \texttt{DynaCI} workflows than baseline-generated workflows.}
\texttt{GH-WCOM} and \texttt{LLM4GW}, i.e., our state-of-the-art baselines, require the most changes (592), with insertions dominating (340) due to two recurring issues. 
First, the tools almost exclusively generate commands that do not implement the functionality of their target. For example, \texttt{GH-WCOM} generates a workflow for \texttt{pallets/flask} that only installs dependencies, e.g., \texttt{pip install -r requirements.txt}, and does not lint the project's source code like the target workflow. 
Second, they default to component configurations that are usually incompatible with their projects. For example, the approaches default to Python 3.8 (or the latest version~\cite{github_2018_setup-python}) when setting up the CI environment, which many projects do not support.

Together, two authors inspected repairs performed with the aforementioned modifications, reporting the repair rate for each tool (\autoref{tab:rq3-breakdown}) as well as their repair category.

\begin{table}[t]
    \centering
    \caption{Repair Rate by Repair Category}
    \resizebox{\columnwidth}{!}{
    \begin{tabular}{l|ccccc}
        \toprule
         \textbf{Repair Category}&  \texttt{\textbf{DynaCI}} & \textbf{\texttt{Heuristics-Only}} &\textbf{\texttt{GH-WCOM}} & \textbf{\texttt{LLM4GW}} & \textbf{\texttt{Commands-Only}}\\
         \midrule
         APT\_PACKAGE\_NOT\_RESOLVED &  5 & 35 &0 & 0 & 48\\
         REFERENCE\_COMMAND\_NOT\_FOUND &  8 & 5 &50 & 50 & 12\\
         SETUP\_UV\_NOT\_FOUND &  1 & 0 &10 & 10 & 9\\
         SETUP\_PYTHON\_NOT\_FOUND &  1 & 1 & 0 & 0 & 40\\
         UNSUPPORTED\_PYTHON\_VERSION &  2 & 5 &37 & 37 & 0\\
         BAD\_ENV\_VARIABLE &  1 & 3 &0 & 0 & 0\\
         \midrule
         ACROSS\_ALL\_CATEGORIES (Repaired / Total Projects)&  22 / 50& 39 / 50&50 / 50& 50 / 50& 50 / 50\\
        \bottomrule
    \end{tabular}}
    \label{tab:rq3-breakdown}
    \vspace{-1em}
\end{table}
\textbf{Observation 7: Substantially fewer \texttt{DynaCI}-generated workflows require repairs than baseline-generated workflows.}
The table shows that among the 50 workflows in our \textit{Evaluation Dataset}, \texttt{DynaCI} and \texttt{Heuristics-Only} workflows only require repairs in 22 and 39 cases, respectively. \texttt{DynaCI} generates 28 out of 50 workflows (56\%) without requiring any manual repair. In contrast, \texttt{GH-WCOM}, \texttt{LLM4GW}, and \texttt{Commands-Only} workflows require repairs in \textit{every} case to add missing functionality. 
We categorize these repairs as follows:

\textbf{APT\_PACKAGE\_NOT\_RESOLVED:} When the name of an \texttt{apt} package cannot be resolved, it cannot be installed. In fact, there are \texttt{apt} repositories, which host different Ubuntu packages, that are registered in the developer's local environment but not in the GitHub Ubuntu runner. 
For example, \texttt{ppa:deadsnakes/ppa}~\cite{deadsnakes_ppa} provides newer Python versions that are unavailable in the default \texttt{apt} repositories in a developer's local environment, but not necessarily~in~GitHub Ubuntu runner. 
As a result, \texttt{Heuristics-Only} may install locally observed \texttt{apt} packages that are unnecessary/unsupported in GHA. \texttt{DynaCI}'s \textit{Refine with LLMs} step mitigates this issue, though developers may still need to remove problematic packages or configure the required repositories.

\textbf{REFERENCE\_COMMAND\_NOT\_FOUND:}
When a GHA workflow job does not run the commands that are executed by the developer in their local environment, it leads to missing functionality in the workflow.
For \texttt{DynaCI} and \texttt{Commands-Only}, this occurs when the LLM removes seemingly extraneous commands (e.g., \texttt{echo}), modifies string literals (e.g., regular expressions in \texttt{grep}), or hallucinates. 
For \texttt{GH-WCOM} and \texttt{LLM4GW}, this arises, in part, because they lack access to developer-executed commands (e.g., \texttt{pytest}, \texttt{tox}, etc.). 
To repair this problem in the generated workflows, developers may manually add the missing commands.

\textbf{SETUP\_UV\_NOT\_FOUND:} When a step cannot find the \texttt{uv}~\cite{github_2024_astral} package manager in its execution environment, its \texttt{uv}-related commands will fail. In our evaluation set, ten workflows use \texttt{uv}, implying that \texttt{GH-WCOM} and \texttt{LLM4GW}, which encountered this problem ten times, failed to install it in each case. To repair this problem, developers may add \texttt{astral-sh/setup-uv}~\cite{github_2024_setup-uv} to failing jobs.

\textbf{SETUP\_PYTHON\_NOT\_FOUND:} 
Python commands will fail if Python is not installed in their execution environment. 
\texttt{Commands-Only} was the only tool that frequently produced this problem by refusing to include actions other than \texttt{actions/checkout} to workflows. 
To repair this problem, developers may add \texttt{actions/setup-python}~\cite{github_2018_setup-python} to failing jobs.

\textbf{UNSUPPORTED\_PYTHON\_VERSION:} 
If a job uses a Python version that is incompatible with its commands, the workflow may fail.  
This issue is common for \texttt{GH-WCOM} and \texttt{LLM4GW}, which install only Python 3.8 or the latest Python 3.x version. 
\texttt{Heuristics-Only} and \texttt{DynaCI} may also encounter this issue when multiple Python versions are observed in the developer's local environment, since they use only the latest version when generating the workflow.
Developers may repair this problem in the generated workflow by specifying a compatible version in \texttt{actions/setup-python}~\cite{github_2018_setup-python}.

\textbf{BAD\_ENV\_VARIABLE:} 
When a job uses an environment variable that causes unexpected behavior, it can lead to workflow failures.
For example, in our local environment, \texttt{mkdocs/mkdocs} \cite{mkdocs} reads the SOURCE\_DATE\_EPOCH variable, which forces builds to use a specified Unix timestamp (e.g., 0 represents January 1, 1970) across their artifacts. While this worked locally, it caused failures in GHA due to its own builds not supporting timestamps before 1980. To repair this problem, developers may manually remove such environment variables that are causing the issues.

%% file: 5-Threats.tex
\section{Threats to Validity}
\label{sec:threats}

\textbf{Internal Validity.} 
To fix errors in the workflows, different reviewers might choose different solutions or modifications, potentially introducing subjectivity or inconsistency in quantifying the required changes. To mitigate this, the first two authors independently analyzed 25 projects and reached a consensus on the number of lines required to make both \texttt{DynaCI}-generated and baseline-generated workflow files functional. 
The first author then applied these criteria to all 50 projects.

\textbf{External Validity.} 
Our experiments use 50 Python projects with GitHub Actions, which may limit generalizability to other languages and CI platforms. However, \texttt{DynaCI} can be adapted to these contexts with modest extensions because it derives workflow components from language-agnostic developer commands and their runtime behavior in local testing environments.
Besides, many CI platforms have similar features~\cite{rostami_2023_ci}; only the components' syntax would need to be modified in \texttt{DynaCI}. 
\texttt{DynaCI}'s divide-and-conquer strategy also enables future heuristics to be added for specific features in other languages. 
While \texttt{DynaCI} currently only supports Linux-based OSes, which is the most common runner in GHA~\cite{bouzenia_2024_gha},  in future, we hope to extend \texttt{DynaCI} to non-Linux environments, including Windows. 
Additionally, to evaluate \texttt{DynaCI}, we needed projects with workflows whose commands could run outside CI, and thus, our results may not fully capture CI-specific features, such as (unofficial) reusable actions. 
However, only 2.05\% of the projects in our \textit{CI Workflow Dataset}, from which our experiment projects were sampled, used such features. 
Our project filtering criteria further ensured that these projects have substantial complexity; thus, these commands are representative of real-world scenarios and projects.

\textbf{Construct Validity.} 
While BLEU and F1 scores capture textual similarity (RQ1), they do not fully account for the correctness of generated workflows in practice. We therefore complement them with a manual analysis of whether generated workflows require modifications to match target executions and how many modifications are needed (RQ3). Together, workflows are evaluated on a textual and a practical level, offering a more comprehensive evaluation of their performance.

%% file: 6-Related.tex
\section{Related Work}
\label{sec:related}
Prior studies have introduced CI workflow generation tools: \texttt{GHAST}~\cite{benedetti_2022_ghast}, \texttt{CIGAR}~\cite{huang_2023_cigar}, \texttt{PACGBI}~\cite{sarschar_2024_pacgbi}, \texttt{GH-WCOM}~\cite{mastropaolo_2024_synthesis}, \texttt{LLM4GW}~\cite{zhang_2024_synthesis}, and \texttt{LLM4CI}~\cite{ghaleb_2025_synthesis}.
Among these tools, the closest to our study are \texttt{GH-WCOM}~\cite{mastropaolo_2024_synthesis} and  \texttt{LLM4GW}~\cite{zhang_2024_synthesis}.
In fact, \texttt{GH-WCOM}~\cite{mastropaolo_2024_synthesis} uses LLMs to complete gaps in a workflow configuration while \texttt{DynaCI} extends the use of LLMs to generate workflows from scratch. Then, \texttt{LLM4GW}~\cite{zhang_2024_synthesis} generates workflows from scratch, but its prompts require domain knowledge about GHA that \texttt{DynaCI} does not. 
Overall, \texttt{DynaCI} reduces developer effort by reusing runtime information from existing local testing environments.

Another line of research focuses on leveraging runtime information for source code generation and analysis~\cite{peng_2025_perfcodegen,wu_2026_environment,chen_2025_runtime}. While \texttt{DynaCI} also uses runtime information, it instead targets CI workflow generation, i.e., configuration synthesis.  
Lian et al.~\cite{lian_2025_validators} show that LLMs are good at validating configurations, such as HBase, without specifically focusing on GHA. 
Those that focused on GHA, for example, Xu et al.~\cite{xu_2025_logsage}, leveraged CI/CD runtime logs to diagnose and repair failures.
However, these approaches do not generate workflows from scratch. To the best of our knowledge, \texttt{DynaCI} is the first runtime-informed approach to CI workflow synthesis.

%% file: 7-Conclusion.tex
\section{Conclusion and Implications}
\label{sec:conclusion}

This paper introduces \texttt{DynaCI}, the first system, to the best of our knowledge, to combine runtime information with LLMs to synthesize GHA workflows. Our evaluation on 50 Python projects shows that \texttt{DynaCI} outperforms state-of-the-art baselines by generating more complete and precise workflows. Below, we distill the implications of our results:

\textbf{Developers may use \texttt{DynaCI} to translate local environments into effective CI workflows with little cost and modification.} Prior work shows that writing CI workflows is challenging~\cite{saroar_2023_gha, zhang_2024_ci, valenzuela-toledo_2022_gha}. Existing tools, such as \texttt{GH-WCOM}~\cite{mastropaolo_2024_synthesis} and \texttt{LLM4GW}~\cite{zhang_2024_synthesis}, still require substantial domain knowledge of GHA. In contrast, \texttt{DynaCI} relies only on the developer's local environment, thus reducing knowledge barriers and effort requirements. Our Observation 1 shows that \texttt{DynaCI} is able to produce more effective workflows than the state-of-the-art tools. 
Observations 6 and 7 show that \texttt{DynaCI}-generated workflows require fewer modifications, while Observation 4 shows that they can be produced at low cost.

\textbf{Tool builders may develop recommendation systems to assist in troubleshooting problems from CI workflow generation tools, including \texttt{DynaCI}, which produced workflows needing the least modification.} CI workflow generation tools can produce workflows that contain problems. While Observations 6 and 7 show that \texttt{DynaCI} workflows require substantially fewer modifications than the state-of-the-art, it can still generate problems, such as missing functionality, that developers would have to correct. Valenzuela-Toledo et al.~\cite{valenzuela-toledo_2022_gha} emphasize the need for tools that could troubleshoot problems in GHA workflows. Tool builders could leverage our repair categories for GHA workflows (\autoref{subsec:evaluation-error}) to build a recommendation system for generated workflow repair.

\textbf{Researchers may adopt \texttt{DynaCI}'s divide-and-conquer approach to generate DevOps configurations beyond CI workflows.} 
Prior work treats CI workflow generation as a single problem, producing an entire workflow at once~\cite{zhang_2024_synthesis, ghaleb_2025_synthesis}.
\texttt{DynaCI} breaks down this problem into \textit{multiple} sub-problems, each targeting an individual component in the workflow; these components are then integrated through a robust refinement process to produce a more effective an efficient workflow that solves the initial problem and requires little modification (Observation 5).
Thus, \texttt{DynaCI} sets the foundation for a family of \textit{divide-and-conquer} approaches to CI workflow generation and beyond.
We encourage researchers to adopt our divide-and-conquer approach to automatically generate other types of DevOps configurations as well (e.g., Docker and Terraform).